\documentclass[intlimits,twoside,a4paper]{article}
\usepackage{amsmath,amssymb}
\usepackage{graphicx,psfrag}
\usepackage[T2A]{fontenc}
\usepackage[cp1251]{inputenc}
\usepackage[eqsecnum]{cmpj2}

\usepackage{slashbox}

\issue{2013}{16}{2}{23602}

\doinumber{10.5488/CMP.16.23602}

\RequirePackage{color}

\title[Phase transitions in the Potts model on complex networks]%
{Phase transitions in the Potts model on complex networks}
\author[M. Krasnytska, B. Berche, Yu. Holovatch]{M. Krasnytska\refaddr{label1,label2}, B. Berche\refaddr{label2}, Yu. Holovatch\refaddr{label1}}

\addresses{
\addr{label1} Institute for Condensed Matter Physics of the National Academy of Sciences of Ukraine,\\
1 Svientsitskii St., 79011 Lviv, Ukraine
\addr{label2}Institut Jean Lamour, Universit\'e de Lorraine, F--54506 Vand\oe uvre les Nancy, France}

\authorcopyright{M. Krasnytska, B. Berche, Yu. Holovatch, 2013}

\date{Received February 14, 2013, in final form March 22, 2013}
\begin{document}
\maketitle
\begin{abstract}
The Potts model is one of the most popular spin models of
statistical physics. {The prevailing majority of work} done so far
corresponds to the lattice version of the model. However, many
natural or man-made systems are much better described by the
topology of a network. We consider the $q$-state Potts model on an
uncorrelated  scale-free network for which the node-degree
distribution manifests a power-law decay governed by the exponent
$\lambda$. We work within the mean-field approximation, since for
systems on random uncorrelated scale-free networks this method is
known  to often give asymptotically exact results. Depending on
particular values of $q$ and $\lambda$ one observes either a
first-order or a second-order phase transition or the system is
ordered at any finite temperature. In a case study, we consider the
limit $q = 1$ (percolation) and find a correspondence between the
magnetic exponents and those describing percolation on a scale-free
network. Interestingly, logarithmic corrections to scaling appear at
$\lambda = 4$ in this case.

\keywords  Potts model, complex networks, percolation, critical exponents
\pacs 64.60.ah, 64.60.aq, 64.60.Bd
\end{abstract}

\section{Introduction} \label{I}
Considerable attention has been paid recently to the analysis of phase
transition peculiarities  on complex networks
\cite{networks_1,networks_2,networks_3,networks_4,networks_5,Dorogovtsev08}. Possible applications of spin models
on complex networks can be found in various segments of physics,
starting from problems of sociophysics \cite{Galam12} to physics of
nanosystems \cite{Tadic05}, where the structure is often much better
described by a network than by geometry of a lattice. In turn,
the Potts model, being of interest also for purely academic reasons,
has numerous realizations, see e.g., \cite{Wu82} for some of them.
The Hamiltonian of the Potts model that we are going to consider in
this paper reads:
\begin{equation}\label{1}
- H=\frac{1}{2}\sum_{i,j}J_{ij}\delta _{n_i,n_j}+\sum _i h_i
\delta_{n_i,0}\,,
  \end{equation}
here, {$n_i=0,1,\dots q-1$, where {$q\geqslant 1$} is the number of
Potts states}, $h_i$ is a local external magnetic field chosen to
favour the $0$-th component of the Potts spin variable $n_i$. The
main difference with respect to the usual lattice Potts Hamiltonian
is that the summation in (\ref{1}) is performed over all pairs $i,j$
of $N$ nodes of {the} network, $J_{ij}$ being proportional to the
elements of  an adjacency matrix of the network. For a given
network, $J_{ij}$ equals  $J$ if nodes $i$ and $j$ are linked and it
equals 0 otherwise.

Being one of possible generalizations of the Ising model, the Potts
model possesses a richer phase diagram. In particular, either first
or second order phase {transitions occur} depending on specific
values of $q$ and $d$ for $d$-dimensional lattice systems
\cite{Wu82}. It is well established by now that this picture is
 {changed} by introducing structural disorder, see e.g.,
\cite{Berche04} and references therein for 2d lattices and
\cite{Berche05} for 3d lattices. Here, we will analyze the impact of
changes in the topology of the underlying structure on
thermodynamics of this model, when Potts spins reside on the nodes
of {an uncorrelated} scale-free network, as explained more in
detail below.

From the mathematical point of view,  the  notion of complex {networks} which has been intensively exploited in
the physical community already for several decades, is nothing else but a {complex graph}
\cite{networks_1,networks_2,networks_3,networks_4,networks_5}. Accordingly, the analysis of the Potts model on various
graphs has already a certain history.  Similar to the Ising model
\cite{zittartz_1,zittartz_2}, the Potts model on a Cayley tree does not exhibit
long-range order \cite{Wang76} but rather a phase transition of
continuous order \cite{Turban80}. Some exact results for the three
state Potts model with competing interactions on the Bethe lattice
are given in \cite{Ganikhodjaev06} and the phase diagram of the
three state Potts model with next nearest neighbour interactions on
the Bethe lattice is discussed in \cite{Ganikhodjaev08}. Potts model
on the Apollonian network {(an undirected graph constructed using
the procedure of recursive subdivision)} was considered in
\cite{Araujo10}. So far, not too much is known about critical
properties of the model (\ref{1}) on scale-free networks of
different types. Two pioneering papers \cite{Igloi02,Dorogovtsev04}
(the latter paper was further elaborated in \cite{Ehrhardt05}) that
used the generalized mean field approach and recurrent relations in
the tree-like approximation, respectively, although agree in
principle about suppression of the first order phase transition in
this model for the fat-tailed node degree distribution, but differ
in the description of the phase diagram. Several MC simulations also
show evidence of the changes in the behavior of the Potts model on a
scale-free network {in comparison with} its 2d counterpart
\cite{Ehrhardt05,Shen11}.  The Potts model on an inhomogeneous
annealed network was considered in \cite{Khajeh07}; relation between
the Potts model with topology-dependent interaction and biased
percolation on scale-free networks was considered in
\cite{Hooyberghs10}. {Smoothing} of the first order phase
transition for the Potts model with large values of $q$ on
scale-free evolving networks was observed in \cite{Karsai07}.

In this paper, we  will calculate thermodynamic functions of the
Potts model on an uncorrelated scale-free network. In contrast to
\cite{Igloi02}, we will work with the free energy (
\cite{Igloi02} dealt with the equation of state). This will enable
us to get a comprehensive list of scaling exponents governing
the second-order phase transition as well as percolation exponents. We
will show the emergence of logarithmic corrections to scaling for
percolation and calculate the logarithmic correction exponents. The
paper is organized as follows. In section \ref{II} we derive general
expressions for the free energy of the Potts model on uncorrelated
scale-free network. Thermodynamic functions will be further analyzed
in section \ref{III}, where we obtain leading scaling exponents and
show the onset of logarithmic corrections to scaling for some
special cases. In section \ref{IV} we will further elaborate the $q=
1$ limit of the Potts model, that corresponds to percolation on a
complex network. We end by conclusions and outlook in section
\ref{V}.

It is our pleasure to contribute by this paper to the Festschrift
dedicated to Mykhajlo Kozlovskii on the occasion of his 60th
birthday and doing so to wish him many more years of fruitful
scientific activity.

\section{Free energy of the Potts model on uncorrelated
scale-free network} \label{II}

In what follows we will use the mean
field approach to analyze the thermodynamics of the Potts model
(\ref{1}) on an uncorrelated scale-free network, that is, a network
that is maximally random under the constraint of a power-law node
degree distribution:
\begin{equation}\label{2}
P(k)=c_\lambda k^{-\lambda},
\end{equation}
where $P(k)$ is the {probability that any given node} has degree
$k$ and $c_\lambda$ can be readily found from the normalization
condition $\sum_{k=k_\star}^{k^\star} P(k)=1$, with $k_\star$ and
$k^\star$ being the minimal and maximal node degree,
correspondingly. For an infinite network, $\lim_{N\to \infty}
k^\star \to \infty$. A model of uncorrelated network with a given
node-degree distribution (called also configuration model, see e.g.,
\cite{Bender78a}) provide a natural generalization of the classical
Erd\"os-R\'enyi random graph and is an undirected graph maximally
random under the constraint that its degree distribution is specified. It has been shown that for such networks the mean field
approach leads in many cases to asymptotically exact results. In
particular, this has been verified for the Ising  model using
recurrence relations \cite{Dorogovtsev02} and replica method
\cite{Leone02} and further applied to $O(m)$-symmetric and
anisotropic cubic models \cite{Palchykov10}, mutually interacting
Ising models \cite{Holyst_1,Holyst_2} as well as to percolation \cite{Igloi02}.
For the Potts model, however, two approximation schemes, the mean
field treatment \cite{Igloi02} and an effective medium Bethe lattice
approach \cite{Dorogovtsev04}, were shown to lead to different
results.

\subsection{General relations}
To define {the} order parameter and to carry out the mean field
approximation in the Hamiltonian (\ref{1}), let us introduce local
thermodynamic averages:
\begin{equation}\label{3a}
\mu_i=\overline{ \delta_{n_i,0}  } \, , \qquad \nu_i=\overline
{ \delta _{n_i,\alpha \neq 0} } \,,
\end{equation}
where the averaging means:
\begin{equation} \label{3b}
\overline{(\dots)} = \frac{{\rm Sp} (\dots )\exp (-H/T)}{{\cal
Z}}\,,
\end{equation}
$T$ is the temperature and we choose units such that the Boltzmann
constant $k_{\rm B}=1$. The partition function
\begin{equation} \label{3c}
{\cal Z} = {\rm Sp} \exp (- H/T),
\end{equation}
and {the} trace is defined by:
\begin{equation} \label{3d}
{\rm Sp} (\dots)= \prod_{i=1}^N \sum_{n_i=0}^{q-1}(\dots).
\end{equation}
 The two quantities defined in (\ref{3a}) can be related using
the normalization condition
$\delta_{n_i,0}+\sum_{\alpha=1}^{q-1}\delta_{n_i,\alpha}=1$, leading
to:
\begin{equation} \label{4}
\nu_i=\frac{(1-\mu_i)}{(q-1)}\,.
\end{equation}
Observing {the} behaviour of averages (\ref{3a}) calculated with
the Hamiltonian (\ref{1}) in the low- and high-temperature limit:
$\mu_i(T\to \infty)=\nu_i(T\to \infty)=1/q$, $\mu_i(T\to 0)=1$,
$\nu_i(T\to 0)=0$ the local order parameter (local magnetization),
$0 \leqslant m_i \leqslant 1$, can be written as:
\begin{equation}\label{5}
m_i=\frac {q\overline{ \delta_{n_i 0}}-1}{q-1}\,.
\end{equation}
Now, neglecting the second-order contributions from the fluctuations
$\delta_{n_i,n_j} - \overline{ \delta_{n_i,n_j} }$ one gets the
Hamiltonian (\ref{1}) in the mean field approximation:
\begin{equation}\label{6}
- H^{\mathrm{mfa}}=\sum_{i,j}J_{ij}\delta
_{n_i,0}m_j+\frac{1}{q}\sum_{i,j}J_{ij}\left[1-m_j+(1-q)m_im_j\right]+\sum _i
h_i \delta_{n_i,0}\,.
\end{equation}
The free energy in the mean field approximation, $-g=
T \ln {\rm Sp} \exp (- H^{\mathrm{mfa}}/T)$, readily follows:
\begin{equation}\label{7}
-g= \frac{1}{q}\sum_{i,j}J_{ij}\left[1-m_j+(1-q)m_im_j\right]+T \sum _i \ln
 \left[ \exp \left( \frac {\sum_j J_{ij} m_j + h_i}{T} \right) + q -1
 \right].
\end{equation}

As usual within the mean field scheme, the free energy (\ref{7})
depends, besides the temperature, both on magnetic field and
magnetization. The latter dependence is eliminated by the free
energy minimization, leading in its turn to the equation of state.
For the Potts model on uncorrelated scale-free {networks} the
equation of state, that follows from (\ref{6}) was analyzed in \cite{Igloi02}. Here, we aim to further analyze temperature and
field dependency of the thermodynamic functions. Contrary to the
mean field approximation for lattice models, where one assumes
homogeneity of the local order parameter (putting $m_i=m$ for
lattices), intrinsic heterogeneity of a network, where different
nodes may have in principle very different degrees, does not allow one
to make such an assumption. One can rather assume within the mean
field approximation that the nodes with the same degree are
characterized by the same magnetization. Therefore, the global order
parameter for spin models on network is introduced via weighted
local order parameters (see e.g. \cite{Holyst13}). Following
\cite{Igloi02} let us define the global order parameter by:
 \begin{equation}\label{8}
m=\frac {\sum_i k_i m_i}{\sum_i k_i}\,.
  \end{equation}
Within the mean field approach, we substitute the matrix
elements $J_{ij}$ in (\ref{7}) by the probability $p_{ij}$ of nodes
$i$, $j$ to be connected. The latter for the uncorrelated network
depends only on the node degrees $k_i$,~$k_j$:
\begin{equation}\label{9}
J_{ij}=Jp_{ij}=J\frac {k_i k_j}{N \langle k \rangle}\,,
\end{equation}
where  $J$ is an interaction constant, $\langle k \rangle
=1/N\sum_{i=1}^N k_i$ is the mean {node degree per
node} \footnote{Such approximation makes the model alike the
Hopfield model used in description of spin glasses and
autoassociative memory \cite{Hopfield_1,Hopfield_2,Hopfield_3,Hopfield_4,Bianconi02,Fernandez09}.}. The
free energy (\ref{7}), being expressed in terms of (\ref{8}),
(\ref{9}), contains sums of unary-type functions over all network
nodes. Using the node degree distribution function, these sums can
be written as sums over node degrees: $\frac{1}{N}\sum_{i=1}^N
f(k_i)=\sum_{k=k_\star}^{k^\star} P(k) f(k)$. In the infinite
network limit, $N\to \infty$, $k^\star\to \infty$, passing from sums
to integrals and assuming homogeneous external magnetic field
$h_i=h$ we get for the free energy of the Potts model on an
uncorrelated scale-free network:
\begin{equation}\label{10}
g=\int_{k_\star}^\infty \left[-\frac{Jk}{q}+\frac{Jk}{q}m+\frac{Jk(q-1)}{q}m^2-T\ln
\left(\re^{\frac{mJk+h}{T}}+q-1\right)\right] P(k)  \rd k \, ,
  \end{equation}
where the node-degree distribution function is given by (\ref{2}).
For small external magnetic field $h$, keeping in (\ref{10}) the lowest
order contributions in $h$, $hm$ and absorbing the $m$-independent terms into the
free energy shift we obtain:
\begin{equation} \label{10b}
 g=\frac{J\langle k \rangle}{q}m+\frac{J\langle k \rangle(q-1)}{q}m^2-T\int_{k_\star}^\infty \ln\left(\re^{mJk/T}+q-1\right) P(k)  \rd k -
\frac{(q-1)\langle k \rangle J}{q^2T}mh\, .
\end{equation}

Free energy (\ref{10b}) is the central expression to be further
analyzed.  In the spirit of the Landau theory, expanding (\ref{10b}) at small
$m$ and first keeping terms $\sim m^2$, one gets for the above
expression at zero external magnetic field:
\begin{equation}\label{11}
 g \simeq -\ln q +\frac{J\langle k \rangle(q-1)}{qT}(T-T_0)m^2 \, ,
  \end{equation}
where $T_0=\frac{J \langle k^2\rangle}{2q\langle k\rangle }$.
Provided that the second moment $\langle k^2 \rangle$ of the
distribution (\ref{2}) exists, one observes that depending on
temperature $T$, the coefficient at $m^2$ changes its sign at $T_0$.
This temperature will be further related to the transition
temperature. Another observation, usual for the spin models on
scale-free networks \cite{Dorogovtsev08} is, that the system remains
ordered at any finite temperature when $\langle k^2 \rangle$
diverges (since $T_0\to \infty$).  For distribution (\ref{2}) this
happens at $\lambda \leqslant 3$. Therefore, we will be primarily
interested in temperature and magnetic field behaviour of the Potts
model at $\lambda > 3$\,\,\,\footnote{Scale-free networks with
$k_\star=1$ do not possess a spanning cluster for $\lambda >
\lambda_{\mathrm{c}}$ ($\lambda_{\mathrm{c}}=4$ for continuous node degree distribution
and $\lambda_{\mathrm{c}}\simeq 3.48$ for the discrete one \cite{Aiello00}). We
can avoid this restriction by a proper choice of $k_\star >1$.}\,. The
expansion of the function under the logarithm in (\ref{10b}) at
small order parameter $m$  involves both the small and the large
values of its argument, $mJk$. To further analyze (\ref{10b}), let
us rewrite it singling out the contribution
(\ref{11})\,\,\footnote{Again, we absorb the constant $-\ln q$ into the
free energy shift.} and introducing {a new integration} variable
$x=mJk/T$:
\begin{equation}\label{12}
 g=\frac{J \langle
k\rangle(q-1)}{qT}(T-T_0)m^2+\frac{c_\lambda
(mJ)^{\lambda-1}}{T^{{\lambda-2}}}\int_{x_\star}^\infty \varphi(x) \rd x -
\frac{(q-1)\langle k \rangle J}{q^2T}mh\, ,
  \end{equation}
  where $x_\star= mJk_\star/T$ and
\begin{equation}\label{13}
\varphi(x)=\left[-\ln \left(\re^{x}+q-1\right)+\ln q
+\frac{x}{q}+\frac{q-1}{2q^2}x^2\right]\frac{1}{x^\lambda}\,.
\end{equation}
Note that the Taylor expansion of the expression in square brackets
in (\ref{13}) at small $x$ starts from $x^{3}$, whereas at large $x$
the function $\varphi(x)$ behaves as $x^{2-\lambda}$ and, therefore,
the integral in (\ref{12}) is bounded at the upper integration
{limit for} $\lambda>3$. To analyze the behaviour of the integral
at the lower integration limit when $m\to 0$ we proceed as follows.

\subsection{Non-integer $\lambda$}

Let us first consider {the case when} $\lambda$ is non-integer.
Then, we represent $\varphi (x)$ for small $x$ as\,\,\footnote{It is
meant in (\ref{14}) and afterwards, that the first sum is equal to
zero if the upper summation limit is smaller than the lower one,
i.e. for $\lambda < 4$.}\,:
\begin{equation}\label{14}
\varphi(x)= \sum_{i=3}^{[\lambda-1]} \frac{a_i}{x^{\lambda-i}}  +
\sum_{i=[\lambda]}^{\infty} \frac{a_i}{x^{\lambda-i}} \, ,
\end{equation}
where $[\ell]$ {is the integer} part of $\ell$,  $a_i\equiv
a_i(q)$ are the coefficients of the Taylor expansion:
\begin{equation}\label{15}
-\ln (\re^{x}+q-1) =\sum_{i=0}^\infty a_i x^i \, .
\end{equation}
The first coefficients are as follows:
\[
a_0=-\ln q, \qquad a_1=\frac{-1}{ q}\,, \qquad a_2=-\frac{q-1}{2q^2}\,,
 \qquad a_3=-\frac{(q-1)(q-2)}{6q^3}\,, \qquad a_4=-\frac{(q-1)(q^2-6q+6)}{24q^4}\,.
 \]
{Integration of the first sum in (\ref{14}) leads to initial}
terms that diverge at $x\to 0$. Let us {extract these} from the
integrand and evaluate the integral in (\ref{12}) as follows:
\begin{equation} \label{16}
\lim_{x_\star \to 0} \int_{x_\star}^\infty  \varphi(x) \rd x = \lim_{x_\star \to 0}
\int_{x_\star}^\infty \left[ \varphi(x) - \sum_{i=3}^{[\lambda-1]}
\frac{a_i}{x^{\lambda-i}} \right] \rd x + \lim_{x_\star \to 0}
\sum_{i=3}^{[\lambda-1]} \int_{x_\star}^\infty \frac{a_i}{x^{\lambda-i}}
\rd x \, .
\end{equation}
For the reasons explained above, the first term in (\ref{16}) does
not diverge at small $m$, { neither does it} diverge at large
$x$, so one can evaluate this integral  at $m=0$ numerically. In what follows we
will denote it as:
\begin{equation} \label{17}
c(q,\lambda) \equiv \int_{0}^\infty \left[ \varphi(x) -
\sum_{i=3}^{[\lambda-1]} \frac{a_i}{x^{\lambda-i}} \right] \rd x \,  .
\end{equation}
Numerical values of $c(q,\lambda)$ at different $q$ and $\lambda$
are given in table~\ref{tab1}.
\begin{table}[h]
\caption{Normalized numerical values of the coefficient
$c(q,\lambda)/(q-1)$, equation~(\ref{17}), for different $q$ and~$\lambda$. \label{tab1}}
\begin{center}
\tabcolsep1.2mm
\begin{tabular}{|c|r|r|r|r|r|r|}
  \hline
  \backslashbox{$\lambda$}{$q$}& $1$ & $2$ & $3$ & $4$ & $6$& $8$  \\
  \hline  \hline
  $5.4$ & $-3.0692$ & $-0.0079$ & $-0.0002$ & 0.0013 & 0.0011& 0.0007  \\
    \hline
  $5.1$ & $-5.9318$ & $-0.0454$ & $-0.0106$ & $-0.0006$ & 0.0028& 0.0025  \\
   \hline
  $4.8$ & $-0.1686$ & 0.0352 & 0.0148 & 0.0058 & 0.0005 & $-0.0005$  \\
   \hline
  $4.5$ & $-0.2439$ & 0.0237 & 0.0154 & 0.0085 & 0.0030 & 0.0012  \\
   \hline
  $4.2$ & $-0.6809$ & 0.0275 & 0.0344& 0.0240 & 0.0119& 0.0067 \\
   \hline
  $3.9$ & 0.5975 & 0.0420 & $-0.0540$ & $-0.0528$ & $-0.0346$ & $-0.0231$  \\
   \hline
  $3.6$ & 0.7240 & 0.0830 & 0.0065 & $-0.0076$ & $-0.0102$ & $-0.0085$  \\
   \hline
  $3.3$ & 1.4001 & 0.2469 & 0.0790 & 0.0315 & 0.0052 & $-0.0010$  \\
  \hline
\end{tabular}
\end{center}
\end{table}

Integration of the second term in (\ref{16}) leads to:
\begin{equation} \label{18}
\sum_{i=3}^{[\lambda-1]} \int_{x_\star}^\infty \frac{a_i}{x^{\lambda-i}}
\rd x =  \sum_{i=3}^{[\lambda-1]} \frac{a_i (x_\star)^{-\lambda
+i+1}}{\lambda-1-i}\, .
\end{equation}
Finally, substituting (\ref{17}) and (\ref{18}) into (\ref{12}) we
arrive at the following expression for the first leading terms of
the free energy at {\em non-integer $\lambda$}:
\begin{eqnarray}\nonumber
 g&=&\frac{J \langle
k\rangle(q-1)}{qT}(T-T_0)m^2+\frac{
c_\lambda c(q,\lambda)}{T^{\lambda-2}}(mJ)^{\lambda-1} +  c_\lambda  \sum_{i=3}^{[\lambda-1]} \frac{a_i
(mJk_\star)^{i}}{\lambda-1-i}T^{1-i} \\ \label{19}
&&{}
-
\frac{J \langle k\rangle (q-1)}{q^2T}mh + O\left(m^{[\lambda]}\right)\, .
 \end{eqnarray}

\subsection{Integer $\lambda$}

Let us consider now the case of  integer $\lambda$. To single out
the logarithmic singularity in  the integral of equation~(\ref{12}), let
us proceed as follows (e.g., see p.~253 in: \cite{Bender78b}). Denoting
 \begin{equation}\label{20}
K(y)=\int_{y}^\infty \varphi(x)\rd x
\end{equation}
we take the derivative with respect to $y$:
\begin{equation}\label{21}
\frac{\rd K(y)}{\rd y} = -\varphi(y) \, .
\end{equation}
Now, $K(y)$ can be obtained expanding the expression in square brackets in
(\ref{13}) at small $y$ and integrating equation~(\ref{21}):
\begin{equation}\label{22}
 K(y) = - \int \varphi(y)  \rd y   = \!\sum_{i=3,i\neq {\lambda-1} }^\infty \frac{a_i y^{i+1-\lambda} }{\lambda-i-1} -a_{\lambda-1}\ln (y)+
 C(q,\lambda),
\end{equation}
with an integration constant $C(q,\lambda)$ and coefficients $a_i$
given by (\ref{15}). Numerical values of $C(q,\lambda)$ at different
$q$ and $\lambda$ are given in table~\ref{tab2}.
 \begin{table} [h]
\caption{$C(q,\lambda)/(q-1)$ for different $q$ and $\lambda$.
\label{tab2}}
\begin{center}
\tabcolsep1.2mm {
  \begin{tabular}{|c|r|r|r|r|r|r|}
    \hline
  \backslashbox{$\lambda$}{$q$} & $1$ & $2$ & $3$ & $4$ & $6$& $8$ \\
     \hline     \hline
    $4$ & $0.9810$  & $0.0355$ & $-0.0085$ & $-0.0134$ & $-0.0109$& $-0.0079$ \\
    \hline
    $5$ & $0.4853$ & $0.0194$ & $-0.0527$& $-0.0483$ & $-0.0303$& $-0.0197$ \\
    \hline
\end{tabular}
}
\end{center}
\end{table}

Substituting $K(mJk_\star)$, cf. equation~(\ref{20}), into (\ref{12}) we arrive at the following expression for
the free energy at {\em integer $\lambda$}:
\begin{eqnarray}\nonumber
 g&=&\frac{J \langle
k\rangle(q-1)}{qT}(T-T_0)m^2 - \frac{c_\lambda
a_{\lambda-1}}{T^{\lambda-2}}(mJ)^{\lambda-1}\ln m+ c_\lambda \left[C(q,\lambda)-
a_{\lambda-1} \ln (Jk_\star/T) \right]\nonumber\\
&&{}\times\frac{(mJ)^{\lambda-1}}{T^{\lambda-2}}+  c_\lambda
\sum_{i=3}^{\lambda-2}  \frac {a_i (mJk_\star)^i }{\lambda-i-1}T^{1-i}-\frac{J \langle k\rangle (q-1)}{q^2T}mh
 + O(m^{[\lambda]})\, .\label{23}
  \end{eqnarray}

Expressions (\ref{19}), (\ref{23}) for the free energy of the Potts model will be analyzed in the
subsequent sections in different regions of $q$ and $\lambda$.

\section{Thermodynamic functions} \label{III}
{Towards an analysis} of the Potts model also in the percolation
limit $q= 1$, let us rescale the free energy by the factor $(q-1)$:
$ g^{'\mathrm{mfa}}= g/(q-1)$ and absorb it by re-defining the free energy
scale. Then, each term in (\ref{19}), (\ref{23}) is also to be
divided by $(q-1)$. Let us use the following notations for several
first coefficients at different powers of $m$ in (\ref{19}),
(\ref{23}):
 \begin{align}\label{25}
A&=\frac{2J\langle k\rangle}{q} \, , \\ \label{26}
B&=-\frac{c_\lambda(Jk_\star)^3(q-2)}{2 q^3(\lambda-4)} \, ,
&B'&=-\frac{c_\lambda J^3(q-2)}{2 q^3} \, , \\ \label{27}
C&=-\frac{c_\lambda(Jk_\star)^4(q^2-6q+6)}{6 q^4(\lambda-5)}
\, ,
&C'&=-\frac{c_\lambda J^4(q^2-6q+6)}{6 q^4} \, , \\
\label{28}
 K&=\frac{c_\lambda J^{\lambda-1} c(q,\lambda)}{(q-1)} \, ,
\\ \label{30}
D&=\frac{J\langle k\rangle}{q^2} \, .
\end{align}
Below, we will start the analysis of thermodynamic properties of the Potts
model by determining its phase diagram in different regions of $q$
and $\lambda$.

\subsection{The phase diagram}\label{IIIa}
To analyze the phase diagram, let us write {down the expressions}
of the free energy at small values of $m$, keeping in (\ref{19}),
(\ref{23}) only the contributions that, on the one hand, allow us to
describe the non-trivial behaviour, and, on the other hand, ensure
thermodynamic stability. Since the coefficients at different powers
of $m$ are functions of $q$ and $\lambda$, cf.
(\ref{25})--(\ref{30}), the form of the free energy will differ for
different $q$ and $\lambda$ as well.

\subsubsection{$1\leqslant q < 2$}\label{IIIa1}
\begin{figure}[ht]
\includegraphics[width=0.45\textwidth]{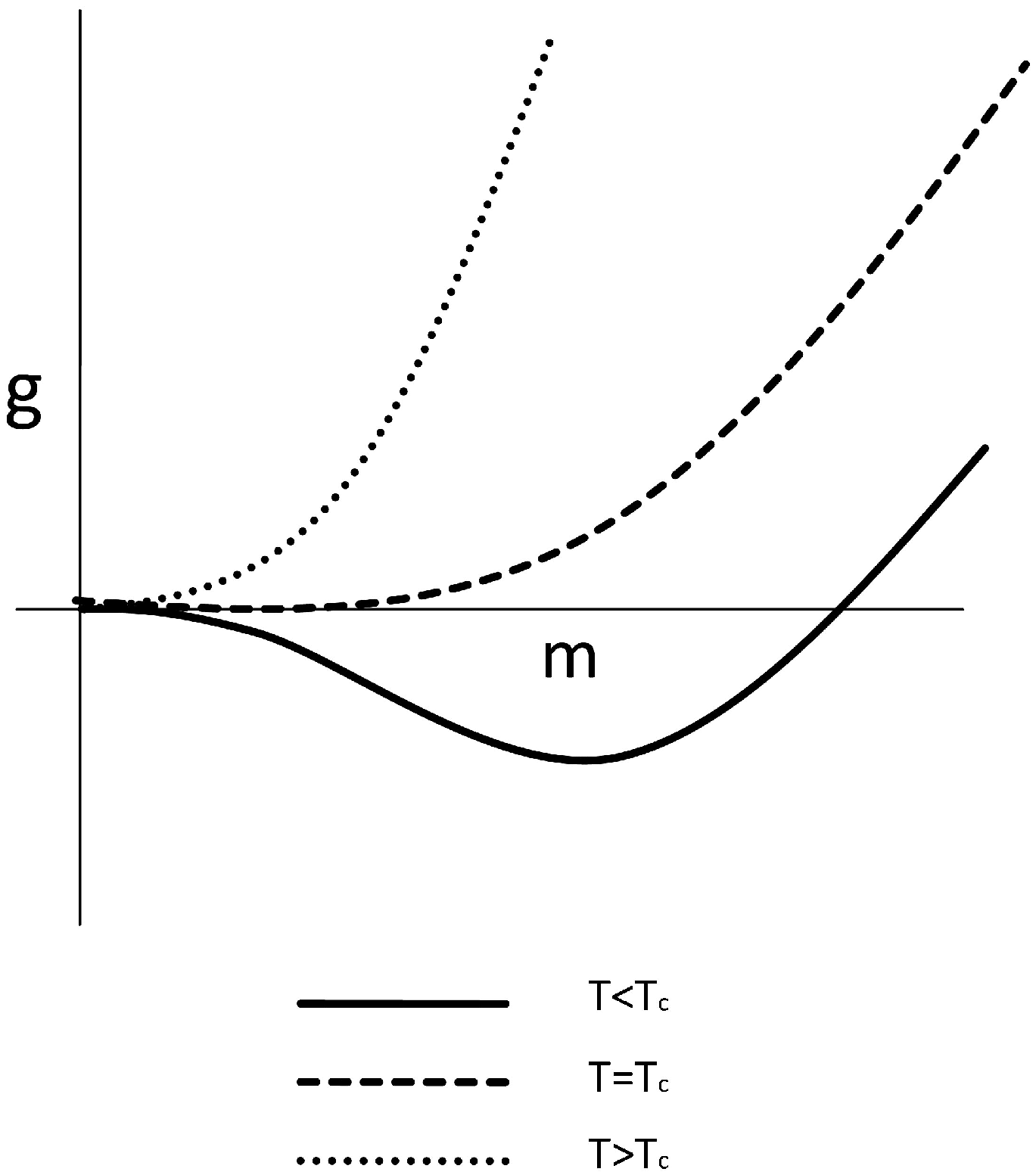}%
\hfill%
\includegraphics[width=0.45\textwidth]{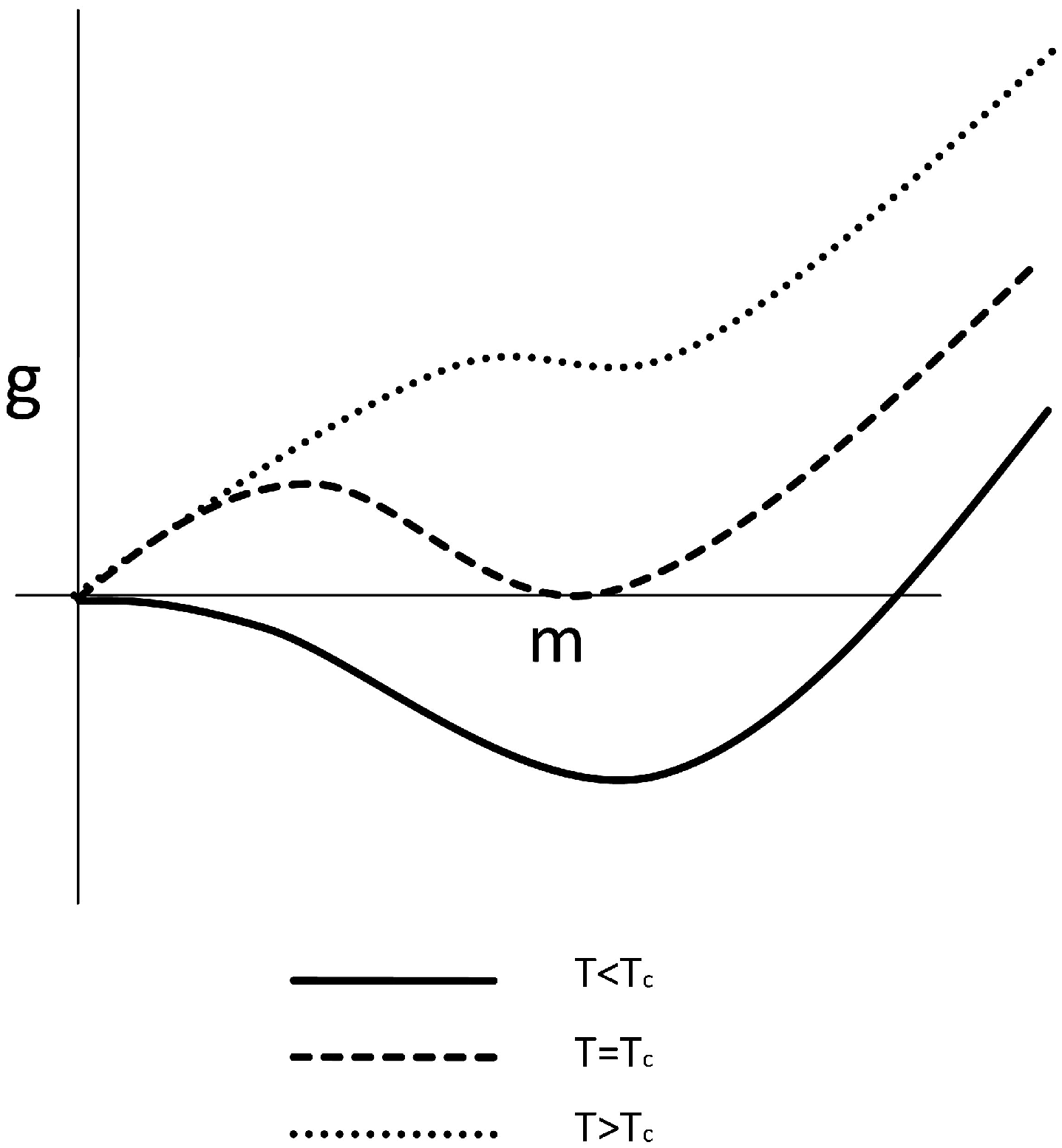}%
\\%
\parbox[t]{0.48\textwidth}{%
\centerline{(a)}%
}%
\hfill%
\parbox[t]{0.48\textwidth}{%
\centerline{(b)}%
}%
\caption{Typical behaviour of the free energy of the Potts model on
uncorrelated scale-free {networks at zero} external field $h=0$.
(a): continuous phase transition; (b): first-order phase
transition.
  \label{fig1}}
\end{figure}

As far as the coefficients $K$, $B$ and $B'$ at $m^{\lambda-1}$,
$m^3$ and $m^3\ln m$ are positive in this region of $q$, it is
sufficient to consider only the three first terms in the free energy
expansion:
\begin{eqnarray}\label{30a}
3< \lambda < 4:& g=&\frac{A}{2T}(T-T_0)m^2+\frac{K}{T^{\lambda-2}} m^{\lambda-1}-\frac{D}{T}mh, \\
\label{30b}
\lambda=4:& g=&\frac{A}{2T}(T-T_0)m^2+\frac{B'}{3T^2}m^3 \ln{\frac{1}{m}} -\frac{D}{T}mh, \\
\label{30c}
\lambda>4:& g=&\frac{A}{2T}(T-T_0)m^2+\frac{B}{3T^2}m^3-\frac{D}{T}mh.
\end{eqnarray}
{The typical} $m$-dependence of functions
(\ref{30a})--(\ref{30c}) at $h=0$ is shown in figure~\ref{fig1}~(a). As it is common for the continuous phase transition scenario,
the free energy has a single minimum (at $m=0$) for $T>T_0$. A
non-zero value of $m$ that minimizes the free energy appears
starting from $T=T_0$. In particular,  the transition remains
continuous in the percolation limit $q=1$, as will be further
considered in sections \ref{IIIb1}, \ref{IV}.

\subsubsection{$q=2$}\label{IIIa2}
For $q=2$, the Potts model corresponds to the Ising model. Indeed,
in this case the coefficient at $m^3$ vanishes and the first terms in
the free energy expansion read:
\begin{eqnarray}\label{31a}
3< \lambda < 5:& g=&\frac{A}{2T}(T-T_0)m^2+K\frac{1}{T^{\lambda-2}}m^{\lambda-1}-\frac{D}{T}mh,
 \\
\label{31b}
\lambda=5:& g=&\frac{A}{2T}(T-T_0)m^2+\frac{C'}{4T^3}m^4\ln {\frac{1}{m}}-\frac{D}{T}mh, \\
\label{31c}
\lambda>5:& g=&\frac{A}{2T}(T-T_0)m^2+\frac{C}{4T^3}m^4-\frac{D}{T}mh.
\end{eqnarray}
It is easy to check that the above coefficients  $K$, $C$, $C'$ are positive for $q=2$. Therefore, again the free
energy behaviour corresponds to the continuous second-order phase transition, see figure~\ref{fig1}~(a).

\subsubsection{$q>2$} \label{IIIa3}
In this region of $q$, phase transition scenario depends on the sign
of the next-leading contribution to the free energy. Indeed, for
positive $K$ the free energy reads:
\begin{equation}\label{32a}
3< \lambda < \lambda_{\mathrm{c}}(q): \hspace{1em} g= \frac{A}{2T}(T-T_0)m^2+K\frac{1}{T^{\lambda-2}}m^{\lambda-1}-\frac{D}{T}mh,
\end{equation}
where $K$ remains positive in the region of $\lambda$ bounded by the marginal value $\lambda_{\mathrm{c}}$ defined by the
condition
\begin{equation}\label{32c}
c(q,\lambda_{\mathrm{c}})=0\, ,
\end{equation}
with $c(q,\lambda)$ given by (\ref{17}).
The free energy (\ref{32a}) is schematically shown in figure~\ref{fig1}~(a) for different $T$. As in the former cases, \ref{IIIa1}, \ref{IIIa2}, it corresponds to the continuous phase transition. With an increase of $\lambda$, the coefficient
$K$ becomes negative and one has to include the next term:
\begin{equation}
\label{32b}
\lambda_{\mathrm{c}}(q)< \lambda < 4: \hspace{1em}  g= \frac{A}{2T}(T-T_0)m^2+K\frac{1}{T^{\lambda-2}}m^{\lambda-1}+\frac{B}{3T^2}m^3-\frac{D}{T}mh,
\end{equation}
$B>0$ for $\lambda <4$. Now, due to the negative sign of the
coefficient at $m^{\lambda-1}$, { the free energy develops a local
minimum for lower $T$ }[see figure~\ref{fig1}~(b)] and {the
order parameter manifests a discontinuity}  at the transition point
$T_{\mathrm{c}}$: scenario, typical of the first order phase transition. With
further increase of $\lambda$, one has {to include more terms} in
the free energy expansion for the sake of thermodynamic stability.
However, the sign at the second lowest order term remains negative,
which corresponds to the free energy behaviour shown in figure~\ref{fig1}~(b): the phase {transition remains  first order}.

\begin{figure}[th]
\centerline{\includegraphics[ width=8cm]{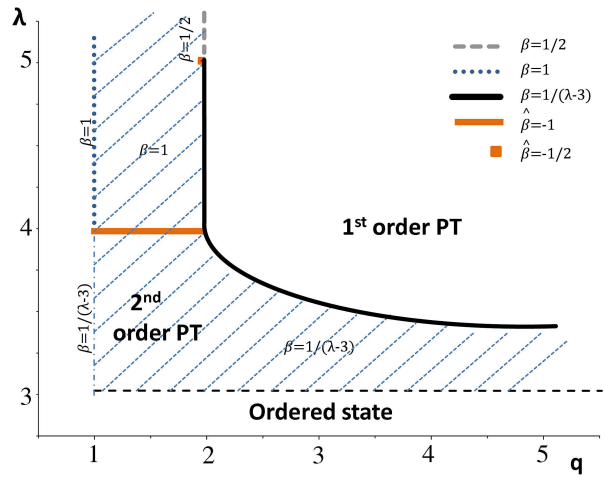}}
\caption{(Color online) The phase diagram of the Potts model {on uncorrelated
scale-free network}. The black solid line separates the 1$^{\rm st}$
order {PT} region from the 2$^{\rm nd}$ order {PT} region
(shaded). The critical exponents along the line are
$\lambda$-dependent. In the 2$^{\rm nd}$ order {PT} region, the
critical exponents are either $\lambda$-dependent (below the light
solid line, red online) or attain the mean field percolation values
(above the red line). For $q=2$, $\lambda \geqslant 5$ (shown by the
gray dashed line), the critical exponents attain the mean field
Ising values. Two different families of the logarithmic corrections
to scaling appear: at $\lambda =5$, $q=2$ (a square, red online) and
at $\lambda =4$, $1\leqslant  q < 2$ (light solid line, red online). For
$\lambda \leqslant 3$ (the region below the black dashed line) the system
remains ordered at any finite temperature. The values for the rest
of critical exponents are listed in tables~\ref{tab3}, \ref{tab4}.
\label{fig2}}
\end{figure}

The above considerations can be summarized in the ``phase diagram"
of the Potts model on uncorrelated scale-free {networks}, that is
shown in figure~\ref{fig2}. Therein, we show the type of the phase
transition for different values of parameters $\lambda$ and $q$.

\subsubsection{{General $q$}, $2< \lambda \leqslant 3$}  \label{IIIa4}
As it was outlined above, for $2 < \lambda \leqslant 3$ the Potts model
remains ordered at any finite temperature. Similar to the Ising
model \cite{Dorogovtsev02}, it is easy to find the high-temperature
decay of the order parameter in this region of $\lambda$ for any
value of $q\geqslant 1$. Since $\langle k^2 \rangle$ becomes divergent
for $2<\lambda \leqslant 3$ {one does not write this term separately
}in the expression for the free energy [cf. (\ref{12})]. As a
result, the corresponding expressions for the free energy read:
\begin{eqnarray}\label{33a}
&& \hspace*{-2em}
2 < \lambda < 3: g=\frac{A}{2T}m^2+
\frac{K'}{T^{\lambda-2}} m^{\lambda-1}-\frac{D}{T}mh,
\\
\nonumber && \hspace*{-2em}
 \lambda = 3: g=\frac{A}{2T}m^2+\frac{c_\lambda J^2}{4q^2 T}m^2 \ln \frac{1}{m}+
 \frac{c_\lambda J^2}{T}\left[ \frac{C'(q,3)}{q-1}+
 \frac{1}{2q^2}\ln\left(\frac{J k_\star}{T}\right)\right]m^{2}-\frac{D}{T}mh\, .
 \\  \label{33b}  &&
\end{eqnarray}
Here, the expressions for the coefficients $K'$, $C'(q,3)$ are the
same as for $K$, $C(q,3)$, equation~(\ref{28}), (\ref{22}) with the only
difference that the function $\varphi (x)$ used for their
calculation does not contain the $x^2$ term. It is easy to check
that the free energies (\ref{33a}), (\ref{33b}) are {minimal for
any finite temperature at a non-zero } value of $m$ that decays at
high $T$ as \cite{Leone02,Igloi02}:
\begin{eqnarray}\label{34a}
2 < \lambda < 3:&& m\sim  T^{-\frac{\lambda-2}{3-\lambda}}\, ,
\\\label{34b}
 \lambda = 3:&&  m\sim  T \re^{-\alpha T},\hspace{1em} \alpha>0\, .
\end{eqnarray}
The above equations (\ref{34a}), (\ref{34b}) give the temperature
behaviour of the mean-field order parameter $m$. The connection with
the magnetization $M$ is found from the self-consistency relation:
 \begin{equation}\label{34}
M=-\Big (\frac{\partial g}{\partial h} \Big )_T\,.
\end{equation}
{One can check }that the solution of this equation at large $T$
is of the form $M\sim \frac{m}{T}$. Correspondingly, this leads to
the following high temperature decay of $M$ \cite{Dorogovtsev02}:
  \begin{eqnarray}\label{34a'}
2 < \lambda < 3:& M\sim & T^{\frac{1}{\lambda-3}}\, ,
\\\label{34b'}
 \lambda = 3:&  M\sim & \re^{-\alpha T},\hspace{1em} \alpha>0\, .
\end{eqnarray}
For the sake of simplicity, in what follows, we will express
the thermodynamic functions in terms of the mean field order
parameter $m$. To get their $M$ dependence, one has to take into
account the above considerations.

\subsection{Regime of the second order phase transition, critical exponents}\label{IIIb}
Let us find the critical exponents {that govern the behaviour}
of thermodynamic functions in the vicinity of the 2nd order phase
transition point $h=0$,$\tau\equiv|T-T_0|/T_0=0$,{where $T_0$ is
the critical temperature of the 2nd order phase transition
($T_{\mathrm{c}}^{\mathrm{2nd}}=T_0$)}. To this end, we will be interested in the
following exponents that govern the temperature and {field dependent
behavior} of the order parameter $m$, the isothermal susceptibility
$\chi_T=(\frac{\partial m}{\partial h})_T$, the specific heat $c_h=T
(\frac{\partial S}{\partial T})_h$, and the magnetocaloric
coefficient $m_T=-T (\frac{\partial S}{\partial h})_T$:
\begin{eqnarray}\label{35a}
 h=0:&&m\sim \tau^\beta , \hspace{0.3cm}  \chi_T\sim \tau^{-\gamma},\hspace{0.3cm} c_h \sim \tau^{-\alpha},
 \hspace{0.3cm} m_T \sim \tau^{-\omega},
 \\\label{35b}
\tau=0:&& m\sim h^{1/\delta}, \hspace{0.3cm}\chi_T\sim
h^{-\gamma_{\mathrm{c}}},\hspace{0.3cm} c_h \sim h^{-\alpha_{\mathrm{c}}},
 \hspace{0.3cm} m_T \sim h^{-\omega_{\mathrm{c}}}.
\end{eqnarray}
Like in the previous subsection, we analyze this behaviour in different
regions of $q$ and $\lambda$. The results of this analysis are
summarized in table~\ref{tab3}.
\begin{table}[tbh]
\caption{Leading critical exponents of the Potts model on
uncorrelated scale-free network. \label{tab3}}
\begin{center}
\tabcolsep1.2mm
\begin{tabular}{|l|l|l|l|l|l|l|l|l|l|}
\hline $q$ & $\lambda$ &$\alpha$ & $\alpha_{\mathrm{c}}$ & $\beta$ &  $\delta$
& $\gamma$ & $\gamma_{\mathrm{c}}$ &
$\omega$ & $\omega_{\mathrm{c}} $ \\
\hline \hline
$\vphantom{\frac{1^1}{1^1}}$
$1\leqslant q\leqslant 2$ & $3<\lambda<4$ & $\frac{\lambda-5}{\lambda-3} $ & $\frac{\lambda-5}{\lambda-2}$ & $ \frac{1}{\lambda-3}$ &  $\lambda-2$ &  $1$ &  $ \frac{\lambda-3}{\lambda-2}$ &  $\frac{\lambda-4}{\lambda-3}$ &  $\frac{\lambda-4}{\lambda-2}$  \\
\hline
$1\leqslant q < 2$  & $\lambda \geqslant 4$ & $-1$    & $-1/2$ &  1 &  2 &   1 &  1/2 &  0 &  0  \\
\hline
$\vphantom{\frac{1^1}{1^1}}$
$q=2$ & $3 < \lambda <5$ & $\frac{\lambda-5}{\lambda-3} $ & $\frac{\lambda-5}{\lambda-2}$ & $ \frac{1}{\lambda-3}$ &  $\lambda-2$ &  $1$ &  $ \frac{\lambda-3}{\lambda-2}$ &  $\frac{\lambda-4}{\lambda-3}$ &  $\frac{\lambda-4}{\lambda-2}$ \\
\hline
$q=2$ & $\lambda \geqslant 5$ & 0    & 0 &  1/2 &  3 &   1 &  2/3 &  1/2 &  1/3  \\
\hline
$\vphantom{\frac{1^1}{1^1}}$
$q>2$ & $3<\lambda\leqslant \lambda_{\mathrm{c}}(q)$ & $\frac{\lambda-5}{\lambda-3} $ & $\frac{\lambda-5}{\lambda-2}$ & $ \frac{1}{\lambda-3}$ &  $\lambda-2$ &  $1$ &  $ \frac{\lambda-3}{\lambda-2}$ &  $\frac{\lambda-4}{\lambda-3}$ &  $\frac{\lambda-4}{\lambda-2}$  \\
\hline
\end{tabular}
\end{center}
\end{table}

\subsubsection{$1\leqslant q<2$} \label{IIIb1}
In this region of $q$, the free energy is given by the expressions
(\ref{30a})--(\ref{30c}). For $T>T_0$, $g$ is {minimal for $h=0$
at} a zero value of the order parameter $m=0$. For $T<T_0$, the
minimum of the free energy corresponds to the non-zero $m$. Based on
the expressions (\ref{30a})--(\ref{30c}) we find in different
regions of $\lambda$:
\begin{eqnarray}\label{36a}
3< \lambda < 4:& m=& T_0^{\frac{\lambda-2}{\lambda-3}}\left[\frac{A\tau}{K(\lambda-1)}\right]^{\frac{1}{\lambda-3}},  \\
\label{36b}
\lambda=4:& m=&\frac{AT_0^2}{B'}\tau|\ln \tau|^{-1}, \\
\label{36c} \lambda>4:& m=&\frac {AT_0^2}{B}\tau.
\end{eqnarray}
{Using formulae} (\ref{36a}), (\ref{36c}) at $q= 1$ we reproduce
the corresponding results for the percolation on scale-free networks
\cite{Cohen02}: the usual mean field percolation result for the
exponent $\beta=1$ for $\lambda>4$ and $\beta=\frac{1}{\lambda-3}$
for $3< \lambda < 4$. Note the appearance of the logarithmic
correction at the marginal value $\lambda=4$. {The resulting
values} of the exponent are given in table~\ref{tab3}. Subsequently,
we obtain {the remaining exponents} defined in (\ref{35a}),
(\ref{35b}) and display them in the first two rows of table~\ref{tab3} as well.

Similar to the order parameter, (\ref{36b}), the temperature and
field behaviour of the rest of thermodynamic functions at
$\lambda=4$ in the vicinity of the critical point is characterized
by the logarithmic corrections. Let us define the corresponding
logarithmic-correction-to-scaling exponents by \cite{Kenna12}:
\begin{eqnarray}\label{37a}
 h=0:\!\!\!&&m\sim \tau^\beta {\mid \ln \tau \mid}^{\hat{\beta}},\quad   \chi_T\sim \tau^{-\gamma} {\mid \ln \tau \mid}^{\hat{\gamma}},\hspace{0.3cm}
 c_h \sim \tau^{-\alpha} {\mid \ln \tau \mid}^{\hat{\alpha}},
\quad  m_T \sim \tau^{-\omega}{\mid \ln \tau \mid}^{\hat{\omega}},
 \\
 \tau=0:\!\!\!&&  m\sim h^{1/\delta} {\mid \ln h \mid}^{\hat{\delta}}, \quad \chi_T\sim h^{-\gamma_{\mathrm{c}}}{\mid \ln h \mid}^{\hat{\gamma_{\mathrm{c}}}},\quad  c_h \sim h^{-\alpha_{\mathrm{c}}}{\mid \ln h \mid}^{\hat{\alpha_{\mathrm{c}}}},
 \quad  m_T \sim h^{-\omega_{\mathrm{c}}}{\mid \ln h \mid}^{\hat{\omega_{\mathrm{c}}}}\, .
 \label{37b}
 \end{eqnarray}
{The obtained corresponding values are given } in table~\ref{tab4}. Note, that all exponents are negative: logarithmic
corrections {enhance the decay} to zero of the decaying
quantities and weaken {the} singularities of the diverging
quantities. We discuss this behaviour more  in detail in section~\ref{IV}.
\begin{table}[tbh]
\caption{Logarithmic-corrections exponents for the Potts model on
uncorellated scale-free network. \label{tab4}}
\begin{center}
\tabcolsep1.2mm
\begin{tabular}{|l||r|r|r|r|r|r|r|r|r|}
\hline
$q$ &$\lambda$ &$\hat{\alpha}$ & $\hat{\alpha_{\mathrm{c}}}$ & $\hat{\beta}$ &  $\hat{\delta}$ & $\hat{\gamma}$ & $\hat{\gamma_{\mathrm{c}}}$ &$\hat{\omega}$ & $\hat{\omega_{\mathrm{c}} }$ \\
\hline\hline
$1\leqslant q<2$ &$\lambda = 4$ & $-2$    & $-3/2$ &  $-1$ &  $-1/2$ &   0 &  $-1/2$ &  $-1$ &  $-1$  \\
\hline
$q=2$ &$\lambda = 5$ & $-1$    & $-1$ &  $-1/2$ &  $-1/3$ &   0 &  $-1/3$ &  $-1/2$ &  $-2/3$ \\
\hline
\end{tabular}
\end{center}
\end{table}

\subsubsection{$q=2$, the Ising  model} \label{IIIb2}
For different $\lambda$, the free energy is given by
(\ref{31a})--(\ref{31c}). Minimizing these expressions one finds for
the order parameter at $h=0$, $T<T_0$:
\begin{eqnarray}\label{38a}
3< \lambda < 5:& m=& T_0^{\frac{\lambda-2}{\lambda-3}}\left[\frac{A\tau}{K(\lambda-1)}\right]
^{\frac{1}{\lambda-3}},  \\
\label{38b}
\lambda=5:& m= & \sqrt{ \frac{AT_0^3}{C'}}\tau^{1/2}|\ln \tau|^{-1/2}, \\
\label{38c} \lambda>5: & m= &\sqrt{ \frac {AT_0^3}{C}}\tau^{1/2}.
\end{eqnarray}
{The corresponding} critical exponents for the other
thermodynamic quantities are given in the third and fourth rows of
table~\ref{tab3}. Logarithmic corrections to scaling (\ref{37a})--(\ref{37b}) appear at $\lambda=5$, their values are given in the
second row of table~\ref{tab4}. Critical behaviour of this model on
an uncorrelated scale-free network was a subject of intensive
analysis, see e.g., the papers
\cite{Leone02,Dorogovtsev02,Palchykov10,vonFerber11} and by the
result given in table~\ref{tab3} we reproduce the results for the
exponents obtained therein.

\subsubsection{$q>2$, $3<\lambda\leqslant \lambda_{\mathrm{c}}(q)$} \label{IIIb3}
In this region of $q,\, \lambda$ the phase transition remains
continuous, see the phase diagram, figure~\ref{fig2}, and the free
energy is given by the expression (\ref{32a}). Correspondingly, one
finds that the spontaneous magnetization behaves as
\begin{equation}\label{39}
3<\lambda\leqslant \lambda_{\mathrm{c}}(q):\,\, m= T_0^{\frac{\lambda-2}{\lambda-3}}\left[\frac{A\tau}{K(\lambda-1)}\right]^{\frac{1}{\lambda-3}}.
\end{equation}
The values of the rest of the critical exponents are given in the
sixth row of table~\ref{tab3}. Since the leading terms of the free
energy (\ref{32a}) at $3<\lambda\leqslant \lambda_{\mathrm{c}}(q)$ coincide with
that of the Ising model at $3<\lambda\leqslant 5$,  (\ref{31a}), the
behaviour of thermodynamic functions in the vicinity of the second
order phase transition is governed by the same set of the critical
exponents: the {Potts model for} $q>2$, $3<\lambda\leqslant \lambda_{\mathrm{c}}(q)$ belongs to the universality class of the Ising model
at $3<\lambda\leqslant 5$. This result was first observed in
\cite{Igloi02} by treating the mean field approximation for the
equation of state.

\subsection{The first order phase transition}\label{IIIc}

For $q>2$, $\lambda > \lambda_{\mathrm{c}}(q)$ the phase transition is of the
first order, see the phase diagram in figure~\ref{fig2}. As we have
outlined in section~\ref{IIIa3}, the next-leading order term of the
free energy has a negative sign and the free energy behaves as shown
in figure~\ref{fig1}~(b). As further analysis shows, the higher the
value of $\lambda$ the more terms one has to take into account in
the free energy expansion in order to ensure the correct $g(m)$
asymptotics. Therefore, in the results given below we restrict
ourselves to the region $\lambda_{\mathrm{c}}(q) < \lambda < 4$, where the free
energy is given by equation~(\ref{32b}). The first order phase transition
temperature {$T_{\mathrm{c}}^{\mathrm{1st}}$} is found from the condition
$g(m=0,{T_{\mathrm{c}}^{\mathrm{1st}}})=g(m\neq 0,{T_{\mathrm{c}}^{\mathrm{1st}}})$, see the red
(middle) curve in figure~\ref{fig1}~(b):
\begin{equation}\label{tc}
{T_{\mathrm{c}}^{\mathrm{1st}}}=T_0+\left[
\frac{-B}{3K(\lambda-1)(\lambda-3)}\right]^{\frac{\lambda-3} {\lambda-4}}\frac{2K(\lambda-1)(\lambda-4)}{A}\,.
\end{equation}
For the jump of the order parameter $\Delta m$ at $T_{\mathrm{c}}$ we find:
\begin{equation}\label{00}
\Delta m=\left[ \frac{-B}{3K(\lambda-1)(\lambda-3)}\right]^{\frac{1}{\lambda-4}}\,.
\end{equation}
Another thermodynamics function to characterize the first order
phase transition is the latent heat $Q$. It is defined by:
\begin{equation}\label{01}
Q= \Delta S \cdot {T_{\mathrm{c}}^{\mathrm{1st}}} \, ,
\end{equation}
where $\Delta S$ is the jump of entropy at $T_{\mathrm{c}}$. With the free
energy given by (\ref{32b}) we find the entropy as
\begin{equation}\label{02}
S=-\Big (\frac {\partial g}{\partial T}\Big)_{h,m}\, .
\end{equation}
Considering the entropy at the transition temperature we can find
the latent heat at the first order phase transition for
$\lambda_{\mathrm{c}}(q)<\lambda<4$:
\begin{equation}\label{69'}
Q=\frac{A}{2}(\Delta m)^2.
\end{equation}

\section{Notes about percolation on scale-free networks} \label{IV}
By the results of section \ref{IIIb1} we  also cover the case $q=1$,
which corresponds to percolation on uncorrelated scale-free networks.
The ``magnetic'' exponents governing corresponding second order phase
transition are given in tables~\ref{tab3}, \ref{tab4}. Let us
discuss them  more in {detail}, in particular relating them to
percolation exponents.  The following exponents are usually
introduced to describe the behavior of different observables near the
percolation\,\,\footnote{For definiteness, let us consider the site
percolation and denote by $p$ here and below  the site occupation
probability.}\, point $p_{\mathrm{c}}$ \cite{percolation_1,percolation_2}: the probability that a
given site belongs to the spanning cluster
\begin{equation}\label{80}
P_\infty \sim (p-p_{\mathrm{c}})^\beta, \qquad  p>p_{\mathrm{c}} \, ,
\end{equation}
the number of clusters of size $s$
\begin{equation}\label{81}
n_s\sim s^{-\tau} \re^{-s/s^*},
\end{equation}
the cluster size at criticality
  \begin{equation}\label{82}
s^*\sim |p-p_{\mathrm{c}}|^{-\sigma},
\end{equation}
the average size of finite clusters
  \begin{equation}\label{83}
\langle s \rangle \sim |p-p_{\mathrm{c}}|^{-\gamma}.
\end{equation}
The above defined exponents $\beta$ and $\gamma$ coincide with the
``magnetic'' exponents $\beta$ and $\gamma$ of the $q=1$ Potts model
(see tables~\ref{tab3}, \ref{tab4}). Therefore, the probability that
a given site belongs to the spanning cluster, and the average size of
finite clusters for percolation on uncorrelated scale-free networks
are governed by the scaling exponents:
\begin{equation}\label{beta}
\beta= \left\{
\begin{array}{cc}
                \frac{1}{\lambda-3}, & \hspace{1cm} 3<\lambda<4, \\
                1 ,& \hspace{1cm} \lambda>4,
              \end{array}
  \right.
\end{equation}
 \begin{equation} \label{gamma}
\gamma=1, \hspace{1cm} \lambda>3.
\end{equation}
The exponents $\tau$ and $\sigma$ may be derived with the help of
familiar scaling relations \cite{percolation_1,percolation_2}:
 \begin{equation}\label{84}
\sigma\beta=\tau-2,
\end{equation}
 \begin{equation}\label{85}
\gamma=\frac{3-\tau}{\sigma}\,.
\end{equation}
Substituting the values of $\beta$ and $\gamma$ (\ref{beta}),
(\ref{gamma}) into (\ref{84}), (\ref{85}) one arrives at the
following expressions for the exponents $\tau$ and $\sigma$:
 \begin{equation}\label{86}
\tau= \left\{
\begin{array}{cc}
                \frac{2\lambda-3}{\lambda-2}, & \hspace{1cm} 3<\lambda<4, \\
                \frac{5}{2} ,& \hspace{1cm} \lambda>4,
              \end{array}
  \right.
\end{equation}
 \begin{equation}\label{87}
\sigma= \left\{
\begin{array}{cc}
                \frac{\lambda-3}{\lambda-2}, & \hspace{1cm} 3<\lambda<4, \\
                1/2 ,& \hspace{1cm} \lambda>4.
              \end{array}
  \right.
\end{equation}
{ Analysing the high-temperature} behaviour of the Potts model
magnetization at $2<\lambda< 3$, (\ref{34a'}), one arrives at the
scaling exponents $\beta=1/(3-\lambda)$, $\gamma=-1$ for the
corresponding observables for percolation at  $p_{\mathrm{c}}=0$.

{Our} formulas (\ref{beta}), (\ref{gamma}), (\ref{86}), and
(\ref{87})  {reproduce} the results for the scaling exponents
that govern percolation on uncorrelated scale-free networks
\cite{Cohen01,Cohen02} as well as those found for the
related models of virus spreading \cite{Pastor01,Moreno01}. All the
above mentioned papers do not  explicitly discuss the case
$\lambda=4$ and possible logarithmic corrections that arise therein.
Moreover, a recent review \cite{Hasegawa11}, that also discusses the
peculiarities of percolation on uncorrelated scale-free networks
does not report on logarithmic corrections [its equation~(95) is
perhaps wrong since it gives no logarithmic corrections for
$\lambda=4$.] {Our results} are in the first row of table~\ref{tab4} where we give a comprehensive list of critical exponents that
govern logarithmic corrections to scaling appearing for the Potts
model at $q=1$, $\lambda=4$, as correctly predicted within the
general Landau theory for systems of arbitrary symmetry on
uncorrelated scale-free networks \cite{Goltsev03}. One may compare
{our values with} the corresponding exponents of  the
$d$-dimensional lattice percolation at $d=6$:
$\hat{\alpha}=\hat{\beta}=\hat{\gamma}=\hat{\delta}=\hat{\alpha}_{\mathrm{c}}=2/7$
(see e.g. \cite{Kenna12}). In this respect, the
logarithmic-correction exponents for the lattice percolation at
$d=6$ and for the scale-free network percolation  at $\lambda=4$
belong to different universality classes. It is easy to check {
that the exponents quoted} in the last  row of {table~\ref{tab4}
 obey} the scaling relations for the logarithmic-corrections
exponents: $\hat{\beta}(\delta-1)=\delta\hat{\delta}-\hat{\gamma}$,
$\hat{\alpha}=2\hat{\beta}-\hat{\gamma}$ \cite{Kenna06_1,Kenna06_2,Kenna06_3},
$\hat{\gamma_{\mathrm{c}}}=\hat{\delta}$,
$\hat{\alpha_{\mathrm{c}}}=\frac{(\gamma+2)(\hat{\beta}-\hat{\gamma})}{\beta+\gamma}+\hat{\gamma}$
\cite{Palchykov10}.

\section{Conclusions and outlook} \label{V}
In this paper we have analyzed the critical behaviour of the
$q$-state Potts model on an uncorrelated scale-free network. The
mean field approach we use in our calculations often leads to
asymptotically exact results when critical behaviour on an
uncorrelated scale-free network is considered. However, in the case
of the Potts model, two similar approximate schemes of calculations,
the mean field \cite{Igloi02} and the recurrent relations for the
tree-like random graphs \cite{Dorogovtsev04} differ in their results
concerning the phase diagram. In particular, for $q>2$ and
$\lambda\leqslant 3$, both approaches predict that the  system always
remains ordered at finite temperature. However, for $\lambda > 3$,
depending on specific values of $\lambda$, the first approach
predicts the first or the second order phase transition, whereas the
second approach predicts the first order phase transition. Our
results complete the above analysis. However, unlike \cite{Igloi02}, where the equation of state was considered, we
have considered the thermodynamic potential which enabled us to
present a comprehensive analysis of temperature and magnetic field
dependence of thermodynamic quantities. It is worth noting
that the model considered here is alike the Hopfield model
\cite{Hopfield_1,Hopfield_2,Hopfield_3,Hopfield_4}, for which the mean field approximation is known to
give exact results.

Our main results are summarized in figure~\ref{fig2} and in tables~\ref{tab3}, \ref{tab4}. Depending on the values of $q$ and on the
node degree distribution exponent $\lambda$, the Potts model
manifests either the first-order or the second-order phase
transition or it is ordered at any finite temperature, see figure~\ref{fig2}. In the second order phase transition region (shaded in the figure),
it belongs either to the universality class of the Ising model on an
uncorrelated scale-free network (with $\lambda$-dependent critical
exponents) or it is governed by the mean field percolation ($1\leqslant q
< 2$, $\lambda \geqslant 4$) or mean field Ising ($q=2$, $\lambda\geqslant 5$)
exponents.

One of the major {points where the critical behaviour of the
Potts model in the second order phase transition regime  differs
from} the Ising {model is that its} logarithmic correction
exponents belong to two different universality classes. As it is
well known, in certain situations, the scaling behaviour is modified
by multiplicative logarithmic corrections (see \cite{Kenna12} for a
recent review and \cite{Berche13} for the specific case of the Potts
model). For lattice systems, such corrections are known to appear,
in particular, at the so-called upper critical dimension, above
which the mean field regime holds. For the scale-free networks,
where the very notion of dimensionality is ill-defined, a change in
the node degree distribution exponent $\lambda$ may turn a system to
such a regime. For the Ising model, this happens at $\lambda=5$ and is
caused by the divergencies in the fourth moment of the node degree
distribution $\langle k^4 \rangle$, as it was analyzed in detail in
\cite{Palchykov10}. In addition to these corrections, for the Potts
model we observe an onset of multiplicative corrections to scaling
at $1 \leqslant q < 2$ for $\lambda=4$. Contrary to the Ising case, these
are caused by the divergencies in the third moment of the node
degree distribution $\langle k^3 \rangle$. This difference in the
{origin of the appearance of these corrections also causes} the
difference in their numerical values: this new set of the
logarithmic correction to scaling exponents belongs to the new
universality class. In particular, they govern percolation on
uncorrelated scale-free networks at $\lambda=4$.

\section*{Acknowledgements} \label{VI}
It is our pleasure to thank Bernat Corominas-Murtra, Yuri Kozitsky,
Volodymyr Tkachuk, and Lo\"ic Turban for useful discussions. This
work was supported in part by the 7th FP, IRSES project N269139
``Dynamics and Cooperative phenomena in complex physical and
biological environments".


\ukrainianpart

\title{Критична поведiнка моделi Поттса на складних мережах}

\author{М.~Красницька\refaddr{label1,label2}, Б.~Берш\refaddr{label2}, Ю.~Головач\refaddr{label1}}

\addresses{
\addr{label1} Інститут фізики конденсованих систем НАН України,\\
вул. Свєнціцького, 1, 79011 Львів, Україна
\addr{label2} Iнститут Ж. Лямура,
Унiверситет Лотарингії, F--54506 Вандевр-ле-Нансі, Францiя}

\makeukrtitle

\begin{abstract}
\tolerance=3000%
Модель Поттса є однією з найпопулярнiших моделей статистичної фiзики.
Бiльшiсть робiт, виконаних ранiше, стосувалась ґраткової версії цiєї
моделi. Однак багато природних та створених людиною систем набагато
краще описуються топологiєю мережi. Ми розглядаємо $q$-станову
модель Поттса на нескорельованiй безмасштабнiй мережi із степенево
згасною функцiєю розподiлу ступенів вузлiв із показником $\lambda$.
Працюємо в наближенні середнього поля, оскiльки для систем на 
нескорельованих безмасштабних мережах цей метод часто дозволяє
отримати асимптотично точнi результати. В залежностi вiд значень $q$
та $\lambda$, спостерiгаємо фазовi переходи першого чи другого роду,
або ж система залишається впорядкованою при будь-якiй температурi.
Також розглядаємо границю $q = 1$ (перколяцiя) та знаходимо
вiдповiднiсть мiж магнiтними критичними показниками та показниками,
що описують перколяцiю на безмастабнiй мережi. Цiкаво, що в цьому
випадку логарифмiчнi поправки до скейлiнгу з'являються при $\lambda=
4$.
\keywords модель Поттса, складнi мережi, перколяцiя, критичнi
показники
\end{abstract}

\end{document}